\documentclass[12pt]{article} 
\usepackage{epsfig}
\usepackage[dvips]{color}
\usepackage[english]{babel}
\title{ TeV scale gravity, mirror universe, and \ldots dinosaurs}
\author{ Z.~K.~Silagadze 
\vspace*{3mm} \\
Budker Institute of Nuclear Physics,  630 090,
Novosibirsk, Russia }
\date{}

\begin{document}
\large
\maketitle

\begin{abstract}
This is somewhat extended version of the talk given at the 
Gran Sasso Summer Institute: Massive Neutrinos in Physics and Astrophysics.
It describes general ideas about mirror world, extra spatial dimensions
and dinosaur extinction. Some suggestions are made how these seemingly 
different things can be related to each other.
\end{abstract}

\newpage

\section{Introduction}
The history of science in particular and the Human history in general
teach us it is not easy to answer a simple question \cite{1}
``What is truth?''. Maybe because the truth usually has infinitely many
aspects or projections to be grasped. The following example from \cite{2}
I like very much. Consider the two figures below.
\begin{figure}[htb]
  \begin{center}
\mbox{\epsfig{figure=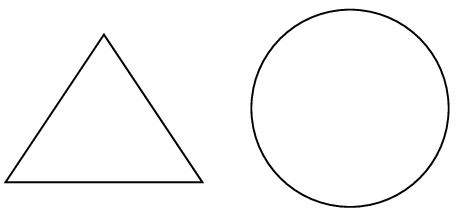
                               ,  height=4.0cm}}
   \end{center}
%\caption {}
\label{Fig1}
\end{figure}

\noindent Are they different? Certainly they are. This seems to be an 
indisputable truth. But if our wonderful divine gift -- our imagination --
will help us to escape bonds of the stiff two-dimensional logic we can see
the following three-dimensional picture:
\begin{figure}[htb]
  \begin{center}
\mbox{\epsfig{figure=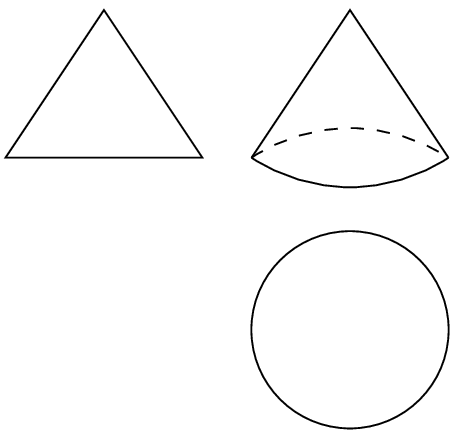
                               ,  height=6.0cm}}
   \end{center}
%\caption {}
\label{Fig2}
\end{figure}

\noindent Now it is clear (an indisputable truth again) that these two 
figures, which originally appeared as two different objects, actually 
are just two different projections of the same thing -- the cone.
From the original two-dimensional picture alone it is impossible to establish
with certainty whether these two figures are substantially different or
not. With the help of imagination we can find as a viable option that these
figures may have the common origin and represent in fact the one essence,
but in order to prove the case one needs further information
(some experiments?)

What follows is an attempt to answer Bludman's question 
\cite{3} ``Mu\ss{} es sein?'' with regard to the Mirror World. Until 
experiments firmly prove or disprove its existence, any answer will
include a great deal of imagination by necessity. So I will describe things
at first sight very different and not related to the Mirror World. I refer
to your imagination to accept a possibility that these different tales are 
in fact fragments of the same story. 

To demonstrate the importance of imagination, I will perform a little
hocus-pocus now and find the Mirror World even in a simple arithmetical
expression.

\section{Arithmetics of the Mirror World}
Let us begin with the (correct) expression
$$ 5+10+1=16.$$
\noindent Is it possible to find the Mirror World in this expression?
Do not be hasty. At least a right handed neutrino and SO(10) GUT can be
found in this innocent expression, as Buccella had reminded us 
recently \cite{4}. But after we catch sight of SO(10) from this 
expression it is possible us to come  across a more advanced 
SO(10)-arithmetics:
$$210+560=770.$$
\noindent The remarkable fact about the fancy numbers above is that all
these numbers are dimensions of some SO(10) irreducible representations
(irreps). Now there is some general problem for you: find all SO(10) 
irreps such that the sum of the first two irrep dimensions matches 
exactly the dimension of the third irrep. 

Maybe after some time you will find this problem a bit tricky and will
decide firstly to try the analogous SO(9)-problem encouraged by the
SO(9)-arithmetical observation
\begin{equation}
44+84=128.
\label{eq1} \end{equation}
\noindent If you are lucky enough, you will find a solution or will 
discover the one, given in the literature by Ramond et al. \cite{5}
and will understand that surprisingly equation (1) has its roots in the
following (simple) triality structure of the $F_4$ exceptional Lie 
algebra:
\begin{figure}[htb]
  \begin{center}
\mbox{\epsfig{figure=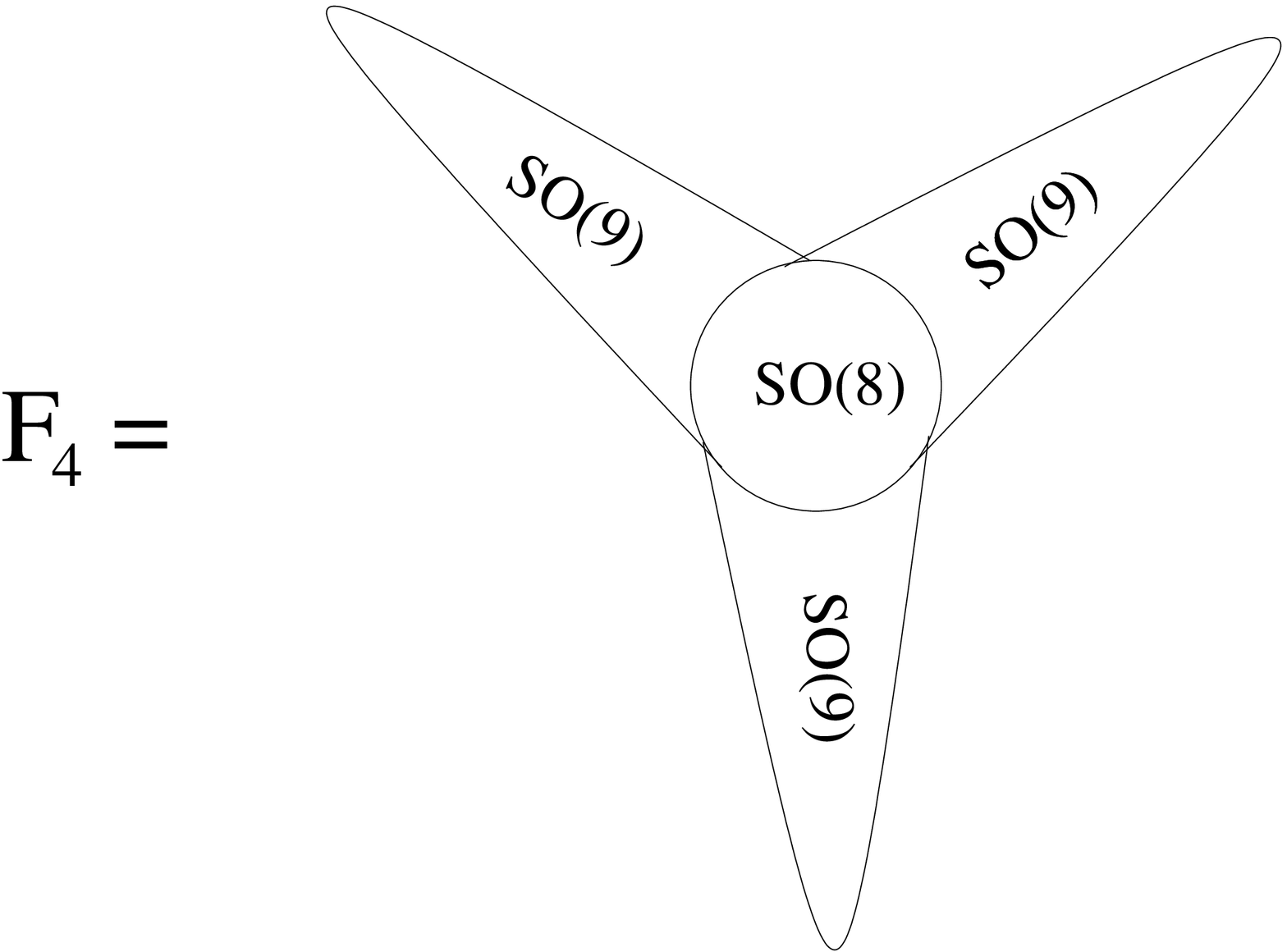
                               ,  height=6.0cm}}
   \end{center}
%\caption {}
\label{Fig3}
\end{figure}

\noindent Meanwhile you will learn a lot of beautiful mathematics like
octonions, triality, Dynkin diagrams, Freudenthal-Tits magic square, Weyl 
chambers etc.

And after you have become so clever, it will strike to you that $SO(9)$ is
nothing but a Wigner's little group associated to the massless degrees of 
freedom of eleven-dimensional supergravity. The irreps from (1) just form
$N=1$ supergravity super-multiplet in eleven dimensions, $\underline{44}$
representing gravitons, $\underline{84}$ -- another massless bosonic field 
and $\underline{128}$ -- the Rarita-Schwinger spinor. So the very equation
(1) ensures supersymmetry, that is the equality between the bosonic and
fermionic degrees of freedom.

But 11-dimensional $N=1$ supergravity is just a low energy limit of much
bigger theory, called $M$-theory \cite{6}. And you will find for sure that
this $M$(arvellous)-theory also gives in various limits all known string
theories in ten dimensions, among them a heterotic string theory which
leads in the low energy limit to the $E_8\times E_8$ effective gauge theory,
this second $E_8$ being nothing but the ``shadow'' world of mirror particles
\cite{6,7,8}!

After I have found the Mirror World even in some simple arithmetical 
expression, you will be not surprised, I hope, to hear that my next topic
is about creatures very closely related to the massive neutrinos. And this
creatures are dinosaurs. 

\section{The dinosaur mystery}
First dinosaurs appeared on the Earth about 250~Myr (million years) ago,
at the beginning of the Paleozoic Era, in a period of time geologists called
``Triassic''. Shortly after their appearance, they grew in size as well as
in numbers and types and dominated the food chain nearly for 200~Myr. Some
dinosaurs were very powerful creatures. Indeed very powerful and very big.
But this did not help them very much when their doomsday came at the end
of the Cretaceous Period, the time period they dominated on the Earth. 
Something
very mysterious happened on the Earth about 65~Myr ago and dinosaurs suddenly
(in a geological time scale) disappeared: their fossils were found throughout
the Mesozoic Era but not in the rock layers of the Cenozoic Era. The first
period of this new Era is called ``Tertiary'' by geologists, so the dinosaur
extinction is known as the Cretaceous--Tertiary or K--T extinction. In fact
dinosaurs were not the only victims of this extinction -- about 85\% of all
species inhabiting the Earth at that time went extinct, among them many
marine species.

Such mass extinctions happened several times in the Earth's history. Let us 
mention some major extinctions \cite{9}:
\begin{itemize}
\item The Precambrian extinction 650~Myr ago -- maybe the first great 
extinction. About seventy percent of the dominant Precambrian flora and 
fauna perished.
\item The Cambrian extinction 500~Myr ago -- about 50\% of all animal 
families went extinct.
\item The Devonian extinction 360~Myr ago -- The crisis primarily affected 
the marine community, having little impact on the terrestrial flora.
\item The Permian extinction 248~Myr ago -- the greatest mass extinction ever
recorded in the Earth's history. About 50\% of all animal families  
perished, as
well as about 95\% of all marine species and many trees. 
\end{itemize}

One can imagine at least two reasons why it is interesting to answer the 
question ``what killed the dinosaurs?'' First of all, without extinctions
we would not be here. Extinction of species is a common companion of 
evolution. A fossil record documents some $2\cdot 10^5$ such extinctions.
Only $\sim 5\%$ of all animal and plant species, ever originated on the 
Earth, are alive today. But this natural extinction process is local and
gradual and do not effect much the evolution. On the contrary, mass 
extinctions are events of global magnitude which nearly destroy the life
on the Earth, but after them the evolution is boosted ahead: new varieties
of species appear which flourish and promptly occupy the vacant
ecological niches. The evolution seems to be a process of punctuated
equilibrium. And this process was certainly punctuated by some global
event 65~Myr ago and as a result the dinosaur era was changed by the
mammal era -- the event clearly of great importance for humankind. But
this is a ``positive'' aspect of mass extinctions. There is a negative one
too. If this unfortunate thing happened to dinosaurs (and many other
less prominent species), there is no guarantee that the same will not
happen to us (humankind) and so it is not excluded that we could be
also found as fossils someday -- the perspective, you certainly do not
like. But ``evolution loves death more than it loves you or me. This is easy 
to write, easy to read, and hard to believe. The words are simple, the 
concept clear-- but you don't believe it, do you? Nor do I. How could I, 
when we're both so lovable? Are my values then so diametrically opposed to 
those that nature preserves? ...we are moral creatures in an amoral world. 
The universe that suckled us is a monster that does not care if we live
or die-- does not care if it itself grinds to a halt. It is fixed and blind, 
a robot programmed to kill. We are free and seeing; we can only try to outwit 
it at every turn to save our skins'' \cite{AnD}. And we can hope to save our 
skins only if we understand where the danger comes from. 

Many theories were suggested to explain the dinosaur mystery. They can
be divided into two general groups \cite{10}. The first kind of
theories operate with the extinction causes which are intrinsic (that
is Earth based) and gradual (last several million years), like
volcanism and plate tectonics.  These are favorite theories of
paleontologists and roughly a half of geologists, attracted by the
problem of dinosaur extinction. Another half of geologists and the
most astronomers and physicists prefer extinction causes which are
extrinsic (of cosmic nature) and sudden, like an asteroid or comet
impact.

The asteroid impact as a cause of the K--T extinction was suggested by
Alvarez et al. \cite{11} and is the most popular hypothesis today.
According to this scenario, the impact of a large object (an asteroid or
a comet with $>10~km$ diameter) 65 Myr ago threw up a huge dust cloud which
remained for weeks and blocked sunlight worldwide. Impact(s) may also
have triggered rounds of volcanic eruptions. As a result, global and
less lasting climate changes, impact-induced global wildfires, acid rains 
etc. effected Earth's ecology of that time enough to force the dinosaurs 
to their end \cite{12}.

The popularity of this hypothesis is based not only on the pagan nature
of the contemporary science. I mean its passion of creating various idols,
and Luis Alvarez was one of such idols in 1980 because of his Nobel prize.  
Simply there is some grave objective evidence that the impact really 
happened at the Cretaceous--Tertiary boundary. The most important evidence
is Iridium anomaly discovered by Alvarez et al. \cite{11}. 

It seems there is a thin band of deposit of clay at the Cretaceous--Tertiary
boundary around the world highly enriched with Iridium. This rare-earth
element is quite sparse in Earth's crust but common in meteorites. So this
Iridium anomaly, which was found by Alvarez et al. initially in marine 
sediments in Italy and afterwards confirmed in both continental and marine
sediments at more than 100 areas world-wide, can be considered as the first
physical evidence that some cosmic intruder hit the Earth 65~Myr ago.

In fact Iridium can be extruded by volcanos from Earth's core where it is
more abundant. And it is known \cite{13} that just about 65~Myr ago India,
which was an isolated island at that time drifting towards its collision 
with Asia, met the head of a mantle plume, molten rock masses
extending from Earth's core-mantle interface upward to the base of
Earth's crust. This mantle plume found its way through India's crust
producing the Deccan Traps volcanism, the greatest volcanic episodes
in the Earth's history ever known.  The hotspot volcano which had produced
Deccan Traps still exists today on Reunion Island and even now is
releasing Iridium \cite{14} !

Therefore one needs some extra evidence to discriminate between impact
and volcano origin of Iridium. These extra evidences are microtektites
(very small glass spheres) strewn fields world-wide and the presence
of quartz grains with multiple sets of shock lamellae (shocked quartz)
in the very same clay layer between Cretaceous and Tertiary
sediments. The both are common products of violent explosions followed
to hypervelocity impacts and
therefore testify in favour of impact, not volcanic origin of Iridium.
The last nail into coffin for competitive theories was the discovery that
the Chicxulub crater located in the Yucatan Peninsula (Mexico) was in fact 
the long sought K--T crater \cite{15}. 

To summarize, there is a little doubt today (especially among astronomers
and physicists) that a large asteroid or comet collided the Earth 65~Myr ago.
It cannot be inferred with certainty that this was the only cause of the K--T
extinction, or even that it was the major cause. Other factors, like Deccan
Traps volcanism, could also play a significant role. Note that the 
competitive ideas suggested to resolve the dinosaur mystery do not necessarily
exclude each other. It may happen that they all contain just different 
projections of the same truth. An interesting example how extraterrestrial and
volcano ideas can be unified is given by Dar \cite{16}. Inspired by the
Hubble Space Telescope discovery that the central star of the Helix Nebula
is surrounded by a ring of about 3500 giant comet-like objects, he 
speculates that similar massive objects can be present in outer solar system.
Gravitational perturbations (for example by passing field stars) can change 
their orbits and bring them into the inner solar system. Near encounter of
the Earth with such ``visiting planet'' can generate gigantic water tidal
waves of $\sim 1~km$ height and crustal tidal waves of $\sim 100~m$ height.
Flexing the Earth by $\sim 100~m$ will release $\sim 10^{34}~ergs$ heat in
Earth's interior in a short time and may trigger the gigantic volcanic
eruptions. Note that the Jupiter's moon Io owes its volcanic activity (the
strongest in the solar system) to the frictional heating due to tidal forces.

But now that's enough about dinosaurs. To proceed and show how dinosaur
extinction is related to massive neutrinos, the main topic of our conference,
we need another mystery story.

\section{The parity mystery}
It is well known that the weak interactions do not respect 
{\bf P}-invariance. To imagine how strange this situation is, let us state
this {\bf P}-noninvariance in another way. The image of our world in a 
{\bf P}-mirror does not look like the original. For example, if we take
15 degrees of freedom of the first quark-lepton generation, after reflection 
in the {\bf P}-mirror we will have (color degrees of freedom is not indicated
for quarks):

\begin{picture}(100,40)
\put(50,27){$u_L \; \; \; d_L \; \; \; e_L \; \; \; \nu_L \; \; \; 
u_R \; \; \; d_R \; \; \; e_R $}
\put(45,19){\line(1,0){183}}
\put(248,17){\bf P}
\put(55,19){\line(-3,-1){10}}
\put(65,19){\line(-3,-1){10}}
\put(75,19){\line(-3,-1){10}}
\put(85,19){\line(-3,-1){10}}
\put(95,19){\line(-3,-1){10}}
\put(105,19){\line(-3,-1){10}}
\put(115,19){\line(-3,-1){10}}
\put(125,19){\line(-3,-1){10}}
\put(135,19){\line(-3,-1){10}}
\put(145,19){\line(-3,-1){10}}
\put(155,19){\line(-3,-1){10}}
\put(165,19){\line(-3,-1){10}}
\put(175,19){\line(-3,-1){10}}
\put(185,19){\line(-3,-1){10}}
\put(195,19){\line(-3,-1){10}}
\put(205,19){\line(-3,-1){10}}
\put(215,19){\line(-3,-1){10}}
\put(225,19){\line(-3,-1){10}}
\put(50,2){$u_R \; \; \; d_R \; \; \; e_R \; \; \; \textcolor{red}{?} 
\; \; \; \; \, u_L \; \; \; d_L \; \; \; e_L $}
\end{picture}

\noindent Therefore we are lacking right-handed neutrino state for the
world to be left-right symmetric! Does this fact mean that the Nature
distinguishes left and right? Not necessarily. In the quantum theory
space inversion is represented by some quantum-mechanical operator 
${\bf P}$.
But different observers can choose not only different conventions about
what is left or right reference frame, but also different bases in the
internal symmetry space of the system. Therefore the operator ${\bf P}$
is determined up to an internal symmetry operator ${\bf S}$. 
In other words, all
operators ${\bf PS_1, \;PS_2, \; PS_3, \ldots}$ are equivalent and any of 
them may be selected as representing space inversion in the Hilbert space
of the quantum system. Now if we find some good enough internal symmetry
$\bf S$, so that ${\bf PS}$ is conserved, the world will be still 
invariant with respect to the ${\bf PS}$-mirror (and this mirror is as 
good as ${\bf P}$-mirror itself for representing space inversion quantum
mechanically). This subtlety in the quantum-mechanical realization of
the space inversion transformation was recognized shortly after the 
experimental discovery of the parity non-conservation and it was suggested
\cite{17} that the charge conjugation ${\bf C}$ could be the very internal
symmetry needed. Indeed the world looks symmetric when reflected in the 
${\bf CP}$-mirror:

\begin{picture}(100,40)(40,0)
\put(50,27){$u_L \; \; \; d_L \; \; \; e_L \; \; \; \nu_L \; \; \; \, 
u_R \; \; \; d_R \; \; \; e_R \; \; \; u_R^C \; \; \; d_R^C \; \; \; 
e_R^C \; \; \;
\nu_R^C \; \; \; u_L^C \; \; \; d_L^C \; \; \; e_L^C$}
\put(45,19){\line(1,0){381}}
\put(445,17){\bf CP}
\put(55,19){\line(-3,-1){10}}
\put(65,19){\line(-3,-1){10}}
\put(75,19){\line(-3,-1){10}}
\put(85,19){\line(-3,-1){10}}
\put(95,19){\line(-3,-1){10}}
\put(105,19){\line(-3,-1){10}}
\put(115,19){\line(-3,-1){10}}
\put(125,19){\line(-3,-1){10}}
\put(135,19){\line(-3,-1){10}}
\put(145,19){\line(-3,-1){10}}
\put(155,19){\line(-3,-1){10}}
\put(165,19){\line(-3,-1){10}}
\put(175,19){\line(-3,-1){10}}
\put(185,19){\line(-3,-1){10}}
\put(195,19){\line(-3,-1){10}}
\put(205,19){\line(-3,-1){10}}
\put(215,19){\line(-3,-1){10}}
\put(225,19){\line(-3,-1){10}}
\put(235,19){\line(-3,-1){10}}
\put(245,19){\line(-3,-1){10}}
\put(255,19){\line(-3,-1){10}}
\put(265,19){\line(-3,-1){10}}
\put(275,19){\line(-3,-1){10}}
\put(285,19){\line(-3,-1){10}}
\put(295,19){\line(-3,-1){10}}
\put(305,19){\line(-3,-1){10}}
\put(315,19){\line(-3,-1){10}}
\put(325,19){\line(-3,-1){10}}
\put(335,19){\line(-3,-1){10}}
\put(345,19){\line(-3,-1){10}}
\put(355,19){\line(-3,-1){10}}
\put(365,19){\line(-3,-1){10}}
\put(375,19){\line(-3,-1){10}}
\put(385,19){\line(-3,-1){10}}
\put(395,19){\line(-3,-1){10}}
\put(405,19){\line(-3,-1){10}}
\put(415,19){\line(-3,-1){10}}
\put(425,19){\line(-3,-1){10}}
\put(50,0){$u_R^C \; \; \; d_R^C \; \; \; e_R^C \; \; \; \nu_R^C \; \; \; 
u_L^C \; \; \; d_L^C \; \; \; e_L^C \; \; \; u_L \; \; \; d_L \; \; \; e_L 
\; \; \;
\nu_L \; \; \; u_R \; \; \; d_R \; \; \; e_R$}
\end{picture}

\noindent Therefore no absolute definitions of left and right are possible 
in the world there ${\bf CP}$ is an unbroken symmetry. 

But we know that in our world ${\bf CP}$ is not an unbroken symmetry. So we 
are left with a strange opportunity that left and right have absolute 
meanings in our world, unless we manage to find some other good internal 
symmetry which will restore the space inversion invariance of the world.
But there is no obvious candidate for such internal symmetry. Therefore the 
scientific community simply became reconciled to the parity non-invariance
of Nature. Moreover, the belief that the only good symmetries are the
proper Poincar\'{e} symmetries became some kind of dogma, as strong as there
was
the opposite belief before Lee and Yang's seminal paper \cite{18} that 
the space inversion and time reversal should also be the exact symmetries
of Nature. This prompt rejection of improper Poincar\'{e} symmetries looks 
especially strange if we remember that an internal symmetry which can
restore the invariance with respect to the full Poincar\'{e} group was in
fact suggested in the very paper \cite{18} of Lee and Yang. Maybe their 
proposal did not gain popularity because at first sight it was no less
strange than the suggestion that the left and right reference frames
are not equivalent. You can restore the equivalence and hence save the
space inversion invariance but you have to pay a price, and the price seems
to be too high: duplication of the world. For any ordinary particle, the
existence of the corresponding ``mirror'' particle is postulated. These
mirror particles are sterile with respect to the ordinary gauge interactions
but interact with their own mirror gauge particles. Vice versa, ordinary
particles are singlets with respect to the mirror gauge group. This mirror
gauge group is an exact copy of the Standard Model 
$G_{WS}=SU(3)_C\otimes SU(2)_L \otimes U(1)_Y$ group with only 
difference that left
and right are interchanged when we go from the ordinary to the mirror 
particles. Therefore the mirror weak interactions reveal an opposite ${\bf
P}$-asymmetry and hence in such an extended universe ${\bf MP}$ is an exact 
symmetry, where ${\bf M}$ interchanges ordinary and mirror particles, and
therefore there is no absolute difference between left and right. This
universe looks symmetric when reflected in the ${\bf MP}$-mirror:
 
\begin{picture}(100,40)(40,0)
\put(50,27){$u_L \; \; \; d_L \; \; \; e_L \; \; \; \nu_L \; \; \; \, 
u_R \; \; \; d_R \; \; \; e_R \; \; \;$ 
\textcolor{magenta}{$u^\prime_R \; \; \;  d^\prime_R \; \; \; 
 e^\prime_R \; \; \;
 \nu^\prime_R \; \; \;  u^\prime_L \; \; \;  d^\prime_L \; \; \; 
 e^\prime_L $}}
\put(45,19){\line(1,0){381}}
\put(445,17){\textcolor{red}{\bf MP}}
\put(55,19){\line(-3,-1){10}}
\put(65,19){\line(-3,-1){10}}
\put(75,19){\line(-3,-1){10}}
\put(85,19){\line(-3,-1){10}}
\put(95,19){\line(-3,-1){10}}
\put(105,19){\line(-3,-1){10}}
\put(115,19){\line(-3,-1){10}}
\put(125,19){\line(-3,-1){10}}
\put(135,19){\line(-3,-1){10}}
\put(145,19){\line(-3,-1){10}}
\put(155,19){\line(-3,-1){10}}
\put(165,19){\line(-3,-1){10}}
\put(175,19){\line(-3,-1){10}}
\put(185,19){\line(-3,-1){10}}
\put(195,19){\line(-3,-1){10}}
\put(205,19){\line(-3,-1){10}}
\put(215,19){\line(-3,-1){10}}
\put(225,19){\line(-3,-1){10}}
\put(235,19){\line(-3,-1){10}}
\put(245,19){\line(-3,-1){10}}
\put(255,19){\line(-3,-1){10}}
\put(265,19){\line(-3,-1){10}}
\put(275,19){\line(-3,-1){10}}
\put(285,19){\line(-3,-1){10}}
\put(295,19){\line(-3,-1){10}}
\put(305,19){\line(-3,-1){10}}
\put(315,19){\line(-3,-1){10}}
\put(325,19){\line(-3,-1){10}}
\put(335,19){\line(-3,-1){10}}
\put(345,19){\line(-3,-1){10}}
\put(355,19){\line(-3,-1){10}}
\put(365,19){\line(-3,-1){10}}
\put(375,19){\line(-3,-1){10}}
\put(385,19){\line(-3,-1){10}}
\put(395,19){\line(-3,-1){10}}
\put(405,19){\line(-3,-1){10}}
\put(415,19){\line(-3,-1){10}}
\put(425,19){\line(-3,-1){10}}
\put(50,0){\textcolor{magenta}{
$ u^\prime_R \; \; \;  d^\prime_R \; \; \;  e^\prime_R \; \; \; 
 \nu^\prime_R \; \; \;  u^\prime_L \; \; \;  d^\prime_L \; \; \; 
 e^\prime_L$} $ \; \; \; u_L \; \; \; d_L \; \; \; e_L 
\; \; \;
\nu_L \; \; \; u_R \; \; \; d_R \; \; \; e_R$}
\end{picture}

After a decade, Kobzarev, Okun and Pomeranchuk returned to this idea
\cite{19}. It was shown that mirror particles should interact only 
extremely weakly with the ordinary particles to evade conflict with 
experiment. In fact only gravity provides a bridge between two worlds.
But gravitational interactions are very weak. So it is not easy to
check the mirror world hypothesis. That's why the idea remained not
popular and even essentially unknown until recently, as illustrated by 
the fact that it was rediscovered by Foot, Lew and Volkas \cite{20}
after another 25 years!

In fact there are also other ways, besides gravity, to connect these 
two worlds.
For example, gauge invariant and renormalizable ordinary-mirror mixing is
allowed for neutral particles like Higgs, $\gamma$ and $Z$ gauge bosons,
and neutrinos.

Higgs -- mirror Higgs mixing can modify significantly the interactions of
the Higgs boson \cite{21}. But we have to wait until the discovery of the
Higgs scalar to test this possibility.

Photon -- mirror photon kinetic mixing term can originate if there exists
mixed form of matter (connector) carrying both ordinary and mirror electric
charges \cite{22}. Even for a very heavy connector, the induced mixing is
expected to be significant and as a result mirror charged particles from
the mirror world acquire a small ($\sim 10^{-3} e$) ordinary electric 
charge. Such millicharged particles have never been found \cite{23}. But
the most stringent bound on the mixing comes from the possibility for 
positronium to oscillate into mirror positronium and disappear \cite{24}.

The neutrino case is the most interesting. Although a possible connection
between neutrino properties and mirror world was noticed earlier \cite{21,
25,26,27}, the real understanding that the mirror world provides a way
to reconcile observed neutrino anomalies (solar neutrino deficit, the
atmospheric neutrino problem, Los Alamos evidence for neutrino oscillations)
arose after two recent papers by Foot and Volkas \cite{28}, Berezhiani and
Mohapatra \cite{29}. The latter work considers an asymmetric mirror world
with spontaneously broken ${\bf MP}$. At present this variant of the mirror
world scenario, further developed in several subsequent publications, is
not excluded by observations. But I will be surprised very much if 
eventually just this asymmetric mirror world proves to be correct. Why, 
just imagine, would God have invented the mirror world if parity remains 
broken?

In the minimal mirror extension of the Standard model, we have just two
neutrino Weyl states $\nu_L$ and $\textcolor{magenta}{\nu^\prime_R}$ 
(mirror particles
are denoted by prime throughout the paper) per generation. If Majorana 
masses are allowed, the most general neutrino mass matrix consistent
with $MP$-parity conservation is \cite{28}
\begin{eqnarray}
\left [ \overline{ \nu_L} \, , \; \overline {(\textcolor{magenta}{
\nu^\prime_R})^C} \right ]
\left ( \begin{array}{cc} M & m \\ m & M^* \end{array} \right )
\left ( \begin{array}{c} (\nu_L)^C \\ \textcolor{magenta}{\nu^\prime_R}
 \end{array} \right ) 
\; + H.c.,
\label{eq2}
\end{eqnarray}
\noindent where the Dirac mass $m$ is real. The mass eigenstates are the
maximal mixtures of ordinary and mirror neutrinos no matter how small the
initial mixing parameter $m$ is:
$$\nu_L^+ =\frac{1}{\sqrt{2}}\left ( \nu_L+(\textcolor{magenta}
{\nu^\prime_R})^C \right ), \;
\nu_L^- =\frac{1}{\sqrt{2}}\left ( \nu_L-(\textcolor{magenta}{\nu^\prime_R}
)^C \right ). $$
\noindent In fact this maximality of mixing is a quite general and very
important consequence of the space inversion symmetry restoration through
mirror world and provides a clear experimental signature of this scenario
\cite{28}.

The mirror world can also naturally accommodate very small neutrino masses
by $MP$-symmetric variant of the standard seesaw model \cite{28}, or it
can even provide an alternative explanation why neutrino masses are so
small \cite{27}. Let us consider the latter case. In order that the neutrino 
not be discriminated as compared to the corresponding charged lepton, let
us assume that in addition to the $\nu_L$ and $\textcolor{magenta}{
\nu^\prime_R}$ states there 
exist a right-handed neutrino $\nu_R$ and its left-handed mirror partner
$\textcolor{magenta}{\nu^\prime_L}$, which are $G_{WS}\otimes G_{WS}$ 
singlets. Such states 
naturally arise if, for example, gauge group of the mirror world 
$G_{WS}\otimes G_{WS}$ is a low energy remnant of $SO(10) \otimes
SO(10)$ grand unification. In such a grand unified mirror world, some early 
stages of symmetry breaking (for example $SO(10) \otimes
SO(10) \to SU(5) \otimes SU(5)$) can generate  a large 
$\nu_R-\textcolor{magenta}{\nu^\prime_L}$ mixing. Besides, ordinary 
electroweak Higgs mechanism 
and its mirror partner will lead to neutrino and mirror neutrino masses. 
Therefore we expect the following neutrino mass terms
\begin{eqnarray}
-{\cal L}_{mass} =  M(\overline{ \nu_R} \textcolor{magenta}{\nu^\prime_L}
+ \overline{\textcolor{magenta}{\nu^\prime_L}}
\nu_R)  + m( \overline{\nu_L} \nu_R+\overline{\nu_R} \nu_L +
\overline{\textcolor{magenta}{\nu^\prime_R}}\textcolor{magenta}
{\nu^\prime_L}+\overline{\textcolor{magenta}{\nu^\prime_L}}
\textcolor{magenta}{\nu^\prime_R}),
\label{eq3} \end{eqnarray}
\noindent where $m$ is expected to be of the order of the charged
lepton mass of the same generation, while the expected
value of M is $10^{14}-10^{15}~GeV$. Among the mass eigenstates 
of (\ref{eq3}) (physical neutrinos denoted by tilde) we have the following
Weyl states 
$$\tilde \nu_L = \cos \theta \, \nu_L- \sin \theta \, \textcolor{magenta}
{\nu^\prime _L} \hspace*{10mm}
\textcolor{magenta}{\tilde \nu^\prime_R} = \cos \theta \, 
\textcolor{magenta}{\nu^\prime_R}-
\sin \theta \, \nu_R \sim {\bf MP}(\tilde \nu_L), $$
\noindent where $\theta \approx m/M$ is very small. These Weyl states 
constitute a very light Dirac neutrino 
$(\tilde \nu _L,\textcolor{magenta}{\tilde\nu ^\prime _R})$ with the 
mass $\sim m^2/M$.
This neutrino is a rather bizarre object -- its left-handed component 
inhabits mostly our ordinary world, while right-handed
component prefers the mirror world intriguing mirror physicists.
Alternatively, you can notice that, because 
${\overline {\textcolor{magenta}{\tilde \nu^\prime}}}_R \tilde \nu_L=
\overline {(\tilde \nu_L)^C} (\textcolor{magenta}{\tilde\nu^\prime_R})^C$,
this ultralight-neutrino mass term
$$m\frac{m}{M}\left(\overline{ \textcolor{magenta}{\tilde\nu^\prime_R}} 
\tilde \nu_L+
\overline {\tilde \nu_L} \textcolor{magenta}{\tilde\nu^\prime_R}\right)$$
\noindent can be considered as a degenerate limit of (\ref{eq2}) with
zero Majorana masses and you can work, if you prefer, in terms of 
(degenerate) maximally mixed ${\bf CMP}$ and mass eigenstates
$$\nu_L^+=\frac{1}{\sqrt{2}}\left ( \tilde \nu_L+
(\textcolor{magenta}{\tilde \nu^\prime_R})^C \right ), \;
\nu_L^- =\frac{1}{\sqrt{2}}\left ( \tilde \nu_L-
(\textcolor{magenta}{\tilde\nu^\prime_R})^C \right ). $$

Besides neutrino oscillations, where are some other observed phenomena
which  can be also interpreted as supporting mirror world hypothesis.
It is well known that there is a lot of dark matter in our universe and
the mirror matter can constitute a considerable fraction of this dark
universe \cite{30}. It is even possible that mirror stars have been
already observed as  gravitational microlensing events \cite{27,31}.
Recent Hubble Space Telescope star counts revealed the deficit of local 
luminous matter \cite{32} predicted by Blinnikov and Khlopov many years
ago \cite{33} as a result of mirror stars existence. Note however that
Hipparchos satellite data \cite{34} have not confirmed the deficit of 
visible matter. Mirror matter was evoked to explain some mysterious
properties of Gamma-ray Bursts \cite{35}. Just during our conference
the paper by Mohapatra, Nussinov and Teplit appeared about the latter 
subject \cite{36}. This paper provokes a thought that maybe the straightest
road from mirror world to the ordinary one lays through extra dimensions.
So we turn our narrative now towards extra dimensions. 

\section{The hierarchy mystery}
The energy scale where gravity becomes strong and quantum gravity effects
are essential is given by the Planck mass. This mass can be estimated as
follows. Suppose two particles of equal masses $m$ are separated at a
distance which equals to the corresponding Compton wavelength $\lambda=1/m$.
If the gravitational interaction energy of the system $G_Nm^2/\lambda=
G_Nm^3$ is of the same order as the particle rest mass $m$, then the
former can not be neglected. This gives for the Plank mass
$$M_{Pl}=\frac{1}{\sqrt{G_N}}\approx 10^{19}~GeV.$$
\noindent Huge difference between this quantum gravity energy scale and 
the electroweak scale $E_{EW}\approx 10^2~GeV$ is astonishing and 
constitutes the so called hierarchy problem. There is also a gauge
hierarchy problem: the Grand Unification scale $E_{GUT}\approx 10^{16}~
GeV$ is very big compared to $E_{EW}$. Any successful theory should not
only explain these hierarchies, but also provide some mechanism to protect
them against radiative corrections. Recently an interesting idea was
suggested by Arkani-Hamed, Dimopoulos and Dvali \cite{37} how to deal
with the hierarchy problem. Certainly, there will be no problem, if there
is no hierarchy. But how can we lower the quantum gravity scale so that the 
hierarchy disappears? It turns out that this is possible if extra spatial 
dimensions exist with big enough compactification radius.  

Suppose besides the usual $x,y,z$ coordinates there exist some additional
spatial coordinates $\textcolor{red}{x_1,\ldots,x_n}$, which are 
compactified on circles
with a common (for simplicity) compactification radius $R$. In such a world
with toroidal compactification, the gravitational potential, created by an
object of mass $m$, should be 
periodic in the extra $n$-dimensions. That is, it should be invariant under 
replacements $\textcolor{red}{x_i} \to \textcolor{red}{x_i} \pm 2\pi R$. 
Besides it should vanish at spatial 
infinity and obey the $(n+3)$-dimensional Laplace equation. These requirements
are satisfied by the following function \cite{38}
\begin{eqnarray}
V=-\sum_{n_1,\ldots,n_n} \frac{\textcolor{red}{\tilde {G}_N} m}{[
r^2+\sum\limits_{i=1}^n (\textcolor{red}{x_i}-2\pi R n_i)^2]^{(n+1)/2}}\; ,
\nonumber \end{eqnarray}
\noindent where  $\textcolor{red}{\tilde {G}_N}$ is the Newton constant for 
$n+4$ space-time dimensions and
$r^2=x^2+y^2+z^2$ is the usual three-dimensional radial distance. If the 
compactification radius $R$ is very large, only the term with 
$n_1=0,\ldots,n_n=0$ survives in the sum and we get the Newton law in 
$n+4$ dimensions:
\begin{equation}
V \approx - \frac{\textcolor{red}{\tilde {G}}_N m}{\tilde{r}^{n+1}},
\label{eq4} \end{equation}
\noindent where $\tilde{r}=\sqrt{r^2+\sum\limits_{i=1}^n \textcolor{red}
{x_i}^2}$.
But if $R\ll r$, the sum can be approximated by an integral
$$V \approx - \frac{\textcolor{red}{\tilde {G}_N} m}
{(2\pi R)^n}\int d^{(n)}\vec{x}
\frac{1}{(r^2+\vec{x}^2)^{(n+1)/2}}\sim -\frac{\textcolor{red}{\tilde {G}_N}}
{R^n}~\frac{m}{r}.$$
\noindent Therefore for the conventional 4-dimensional Newton constant we
have
$$G_N \sim \frac{\textcolor{red}{\tilde {G}_N}}{R^n}.$$
\noindent On the other hand, the fundamental multidimensional quantum
gravity scale $\textcolor{red}{\tilde M_{Pl}}$ is now determined from
$$|\textcolor{red}{\tilde M_{Pl}}V(\frac{1}{\textcolor{red}{\tilde M_{Pl}}})
|\sim \textcolor{red}{\tilde M_{Pl}},$$
\noindent where the potential $V$ is given by the equation (\ref{eq4}),
and we have
$$\textcolor{red}{\tilde M_{Pl}}=\left [ \textcolor{red}{\tilde G_N}
\right ]^{-\frac{1}{n+2}}.$$
\noindent The last two relations indicate
\begin{eqnarray}
\frac{M_{Pl}}{\textcolor{red}{\tilde M_{Pl}}}\sim 
\left(\frac{R}{\textcolor{red}{R_0}}\right)^{\frac{n}{2}} \; ,
\label{eq5}\end{eqnarray}
\noindent where $\textcolor{red}{R_0}=1/\textcolor{red}{\tilde M_{Pl}}
$ and $\textcolor{red}{R_0}\approx 10^{-19}~{\mathrm m}$ (m -- one meter),
if the fundamental quantum gravity scale $\textcolor{red}{\tilde M_{Pl}}$ 
is in a few TeV
range. Therefore the initial $M_{Pl}/E_{EW}$ hierarchy problem can be
traded to another hierarchy: the largeness of the compactification radius
compared to $\textcolor{red}{R_0}$. Namely, we get from (\ref{eq5}) 
the corresponding compactification radius as
$$R\sim 10^{\frac{32}{n}-19}~{\mathrm m}.$$
\noindent For one extra dimension this means modification of the Newton's
gravity at scales $R=10^{13}~{\mathrm m}$ and is certainly excluded. But 
already for $n=2$, $R\sim 1~{\mathrm mm}$ -- just the scale where our 
present day experimental knowledge about gravity ends.

Although gravity was not checked in the sub-millimeter range, Standard
Model interactions were fairly well investigated far below this scale.
Therefore if the large extra dimensions really exist, one needs some
mechanism to prevent Standard Model particles to feel these extra 
dimensions. Remarkably, there are several possibilities to ensure their
confinement at a 3-dimensional wall in the multidimensional
space \cite{39}. Just to illustrate one of them, let us consider a toy 
model \cite{40,41} in the (3+1)-dimensional space-time with the Lagrangian
\begin{eqnarray}
{\cal L}=\bar \psi i\hat \partial \psi -h\phi \bar \psi \psi+
\frac{1}{2} (\partial_\mu \phi)^2-\lambda(\phi^2-v^2)^2.
\label{eq6} \end{eqnarray} 
\noindent This Lagrangian possesses $Z_2$ symmetry
$$ \psi \to i\gamma_5 \psi, \; \; \phi \to -\phi, $$
\noindent which is spontaneously broken in the true vacuum state where
$<\phi>=v$ or $<\phi>=-v$. We assume that the spinor-scalar interaction
term $h\phi \bar \psi \psi$ is small, so in a good approximation the equation
of motion for the field $\phi$ looks like
\begin{eqnarray}
\partial_\mu \partial^\mu\phi=-4\lambda\phi(\phi^2-v^2).
\label{eq7} \end{eqnarray}
\noindent It is easy to check that (\ref{eq7}) has a kink-like solution which
depends only on the $z$-coordinate
$$\tilde \phi(z)=v\tanh{(\sqrt{2\lambda}vz)}.$$
\noindent This solution is a domain wall interpolating between two different
vacua $<\phi>=v$ and $<\phi>=-v$. Its thickness in the $z$ direction is of 
order of $m^{-1}$, where $m=\sqrt{2\lambda}v$.

Let us consider now the fermion in this kink-like background. The equation
of motion which follows from(\ref{eq6}) is
\begin{eqnarray}
i\hat\partial \psi=h\tilde \phi(z)\psi.
\label{eq8}\end{eqnarray}
\noindent This last equation has a factorized solution
$$\psi=\nu (x,y) f(z),$$
\noindent where $f(z)$ is a scalar function and the $\nu (x,y)$ spinor
satisfies (note that $\gamma_3$ is anti-hermitian)
$$i\hat\partial\nu (x,y)=0, \; \; \gamma_3 \nu (x,y)=i\nu (x,y).$$
\noindent For $f(z)$, equation (\ref{eq8}) then gives
$$\frac{df(z)}{dz}=-h\tilde \phi(z)f(z),$$
\noindent its solution with $f(0)=1$ being
$$f(z)=\exp{ \left \{-h\int_0^z \tilde \phi(z)dz\right \}}=
\exp{ \left \{-\frac{h}{\sqrt{2\lambda}}\ln{(\cosh{zm})}\right \}}.$$
\noindent We see that
$$\psi=\nu (x,y) \exp{ \left \{-\frac{h}{\sqrt{2\lambda}}
\ln{(\cosh{zm})}\right \}}$$
\noindent describes a massless ``flat'' fermion $\nu (x,y)$ localized on the
domain wall, the localization scale determined by the fermion-scalar
interaction strength $h$.  

To summarize, the hierarchy mystery maybe indicates the following fascinating
structure of our world: the Standard Model particles (and hence human 
observers) are stuck on a wall (``3-brane'') in the higher 
($\ge 4+n$) dimensional space-time. On the contrary, gravity propagates
freely in the remaining space (the bulk) and feels large ($\sim 1~{\mathrm
mm}$) compact extra dimensions. In the string theory framework, this picture
is naturally achieved if the ordinary particles correspond to
endpoints of open strings attached to the brane, while gravity,
represented by closed strings, can propagate in the bulk. The most surprising
thing about this crazy idea is that it doesn't come in immediate conflict
with known experimental facts \cite{39,Ant,42}.

I'd like to end this chapter with some my personal experience with extra
dimensions. Some times ago I had sent an e-mail letter to my friend in Chicago.
Soon I received the answer saying ``I have received a message from you but I 
don't know to which Sasha it is addressed (I don't know about any Sasha now 
in Milano)''. I was surprised, not so much by what my letter went to Milan 
instead of Chicago, but by the fact that the answer was from Andrea Gamba,
and while preparing my diploma theses at my university years I had read 
a very interesting paper by A.~Gamba \cite{43} about peculiarities of the
eight-dimensional space. I was intrigued and asked him if he was the very
eight-dimensional Gamba. The answer was ``It's really a mystery how I 
received your letter; unfortunately I don't know
about 8-dimensional space, in 1967 I was 5 years old... But certainly your 
message passed through some extra dimension!'' 

So personally I'm quite convinced about existence of extra dimensions. I was
so much astonished by the coincidence described above that I even wrote a 
scientific paper \cite{44} about peculiarities of the eight-dimensional space 
and its possible connection to the generation problem -- this paper can be
considered as a material evidence of communications through extra dimensions.
But now it's time to stop making fun and ask what profit the large extra 
dimensions can give for the mirror world.

\section{Extra dimensions and the mirror universe}
Gravity is the main connector between our and mirror worlds. Therefore, if it 
becomes strong at high energies of about few TeV, the immediate consequence
will be a possibility to produce mirror particles at future high energy 
colliders via virtual graviton exchange. The typical total cross-sections are
\cite{45}
$$\sigma \sim \frac{s^3}{\Lambda^8}\sim ({\mathrm few \; \; \; pb})~
\left (\frac{s}{{\mathrm TeV}^2}\right)^3
\left (\frac{{\mathrm TeV}}{\Lambda}\right )^8, $$
\noindent where $\Lambda \sim 1 {\mathrm TeV}$ is an ultraviolet cutoff 
energy for the effective low-energy theory, presumably of the order of 
the bulk Planck mass \cite{46}. These cross sections are quite sizeable,
but unfortunately there is no clear experimental signature for such
kind of events. May be more useful signature have reactions accompanied by
the initial-state radiation but we expect severe background problems here, 
in particular from the real graviton emission. Therefore the TeV-scale
quantum gravity can allow quite effective mirror matter production at
future TeV-range colliders, but it will be very difficult to convince
skeptics that the mirror particles have been really produced.

Another interesting effect is quarkonium -- mirror quarkonium oscillations.
As a result, heavy C-even quarkonia can 
oscillate into their mirror counterparts, and hence disappear from our world.
Unfortunately the expected probabilities are  very small \cite{45}. For
example, the probability for $\chi_{b2}$ state to oscillate into its mirror 
partner is about $3\cdot 10^{-14}$.

The most promising effect is connected to mirror supernova, because some 
part of a mirror supernova energy will be released in our world too.
In \cite{45} $\textcolor{magenta}{ e^{\prime +} e^{\prime-}}
\to e^+ e^-,\; \gamma \gamma $ reactions were considered as a tool to 
transfer energy from the mirror to the ordinary sector. The resulting ordinary 
energy emissivity per unit volume per unit time of a mirror
supernova core with a temperature $T$ is given by the thermal average over the
Fermi-Dirac distribution and was found to be \cite{45}
\begin{eqnarray}
\dot{q}=\frac{6T^{13}}{25\pi^3\Lambda^8}[I_5(\nu)I_6(-\nu)+
I_5(-\nu)I_6(\nu)],
\label{eq9} \end{eqnarray}
\noindent where
$$ \nu=\frac{\mu_e}{T} \; \; \; {\mathrm and} \; \; \;
I_n(\nu)=\int\limits_0^\infty dx \frac{x^n}{\exp{(x+\nu)}+1}, $$
\noindent $\mu_e$ being the chemical potential for mirror electrons in the
mirror-supernova core.

Let us compare (\ref{eq9}) to the neutrino emissivity by
supernova \cite{47} (only the leading term is shown)
\begin{equation}
\dot{q}_{\nu \bar \nu}=\frac{2G_F^2T^{9}}{9\pi^5}(C_V^2+C_A^2)
[I_3(\nu)I_4(-\nu)+I_3(-\nu)I_4(\nu)],
\label{eq10} \end{equation}
\noindent where $C_A=\frac{1}{2}, \; C_V=\frac{1}{2}+2\sin^2{\Theta_W}$ and
$G_F$ is the Fermi coupling constant. For the core temperature 
$T=30~{\mathrm MeV}$, chemical potential $\mu_e\approx 345~{\mathrm MeV}$ 
and $\Lambda \sim 1~{\mathrm TeV}$, the last equations (\ref{eq9}) and
(\ref{eq10}) give
$$\frac{\dot{q}}{\dot{q}_{\nu \bar \nu}}\approx 1.4\cdot 10^{-16}.$$
\noindent As expected, we get a very small number. But in the first 
$\sim 10$ seconds the neutrino luminosity from
a supernova is enormous \cite{48}: $L_{\nu \bar \nu}\approx 3\cdot 10^{45}
W$ for each species of neutrino. And even $1.4\cdot 10^{-16}$-th part of
$L_{\nu \bar \nu}$ is thousand times larger than the solar luminosity!

Therefore mirror supernovas can be seen by ordinary observers, at least
for some seconds after their birth. Note that according to \cite{36} we are 
already observing light from mirror supernovas as gamma ray bursts!

We tacitly assumed above that the ordinary and mirror matter are located
on the same 3-brane. For space-times with extra dimensions this is not
necessarily the only possibility. In fact you can imagine a situation
then different worlds are located on different 3-branes \cite{49} (or even
on branes with dimensionality other than 3). But I would be careful of
using the nickname ``mirror'' for particles living on different brane.
Maybe ``shadow world'' or ``parallel world'' is more appropriate in this 
case. I prefer to
reserve the name ``mirror world'' for situations which mean the exact
parity symmetry. But how the exact parity invariance can be reconciled
with parallel worlds? A priori one can't expect any symmetry between
parallel worlds which are located on different branes. For me the only
natural possibility is to ensure the parity symmetry for separate brane
worlds. I think this may be achieved if particles can't cross the brane
(in the low energy approximation) and are trapped on the different surfaces
of the brane. Then the parity transformation will involve a transition from
one brane surface to another. Therefore the mirror particles are just
particles located on the another surface of our brane and so are not 
separated from the ordinary world very much in extra dimension, if the brane
is thin. In this case one should expect the same low scale quantum gravity
effects as discussed at the beginning of this chapter for the situation then
the ordinary and mirror particles inhabit the same brane.

This idea is not as wild as it seems at first sight. Let me recall you an
interesting condensed matter analogy: vierbein domain walls in superfluid 
$^3He$-$A$ film \cite{50}. Such domain wall divides the bulk into two 
classically separated ``worlds'': no quasiparticle can cross the wall in the
classical limit. But ``Planck scale physics'' allows these worlds to 
communicate and quasiparticles with high enough energy can cross the wall.
Moreover, the left-handed chiral quasiparticle becomes right-handed when the
wall is crossed!

\begin{figure}[htb]
  \begin{center} \mbox{\epsfig{figure=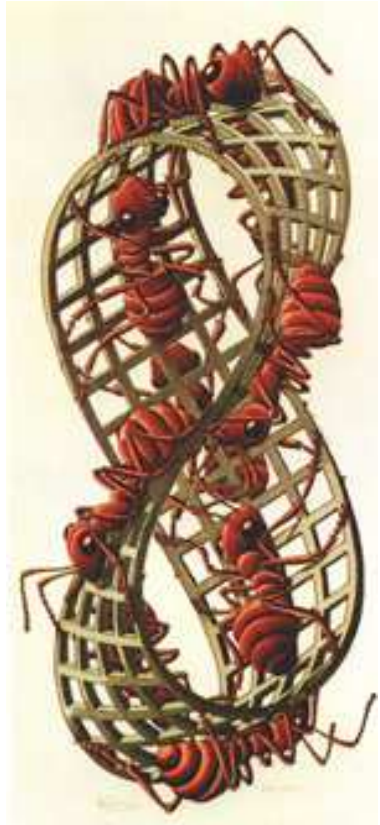 , height=11.0cm}}
  \end{center}
  \caption {The M\"{o}bius world.}
  \label{Fig4}
\end{figure}

If you want a really cool crazy idea -- here it is: the mirror world without
mirror particles \cite{51}. To illustrate this idea, imagine you are the king
of ants living in a two-dimensional flatland. One day your main court 
astrologist gives you a piece of exciting news that there is a deep sense in 
the notions of left and right, because nature does not respect parity 
symmetry and so the absolute meaning of the left, as the side preferred by
stars, can be established. You immediately decide to notify your subject ants
to what is left -- the lucky side. So you send couriers with this mission 
throughout your kingdom. It may happen however that your world has a 
non-trivial global structure in the higher dimensional space and constitutes,
for example, a  M\"{o}bius strip. Then after some time one of your 
couriers can 
be found in a land, your main astrologist calls the land of shadows. You can 
not see him but can communicate with him using gravity. Gravitationally you 
feel as if he were somewhere very close. And really he is just beneath you 
on the M\"{o}bius strip -- see Fig.\ref{Fig4} below \cite{52}. 

But you are flat, as are all of your subjects, and so have no idea about
extra dimensions. You can't say that your courier ant is turned upside-down,
because he is two-dimensional. And his two-dimensional appearance, checked
by gravity, looks the same as for all other ants. Simply in his zeal to
fulfill your order he traveled too far away. And everybody knows in your 
kingdom that if you travel long enough way you will return the same place,
but will return as an invisible shadow. Your main astrologist says that one 
can reach the land of shadows after very long journey. But anyway this 
land of 
shadows is a part of your kingdom -- nobody, even your main astrologist, 
can tell you where the ordinary land ends and the land of shadows begins. 
So 
naturally you want your shadow subjects also to have the correct notion
to what is the left side. And here a great surprise is awaiting you. For
your main astrologist horror, you shadow courier indicates completely
different side as the left side -- the side which originally was marked as 
right by the very same courier before he left the court.

Hence in a such M\"{o}bius world the absolute difference between left and 
right has meaning only  locally. No such difference can be established 
globally -- the world as the whole is parity invariant!

If you do not like worlds to have edges, you can consider, for example,
a Klein's bottle universe instead. In this case you need at least four
space dimensions to realize such (two-dimensional) world without 
self-intersections.

\section{Nemesis -- the dark (matter) sun?}
But, for goodness's sake, what have in common all these mirror worlds and
extra dimensions with dinosaurs? -- you may ask. To explain this, we need
one more (in fact my favorite) dinosaur extinction theory \cite{53}:

``There is another Sun in the sky, a Demon Sun we cannot see. Long ago, 
even before great grandmother's time, the Demon Sun attacked our Sun. 
Comets fell, and a terrible winter overtook the Earth. Almost all life was 
destroyed. The Demon Sun has attacked many times before. It will attack 
again.'' 

It is a very nice theory, having almost mythical power, isn't it? But such
explanation would be enough in some primitive society, not spoiled by the 
science and civilization. You need more scientific story, I suspect. And the 
scientific story begins with the question: are mass extinctions periodic?

\begin{figure}[htb]
  \begin{center} \mbox{\epsfig{figure=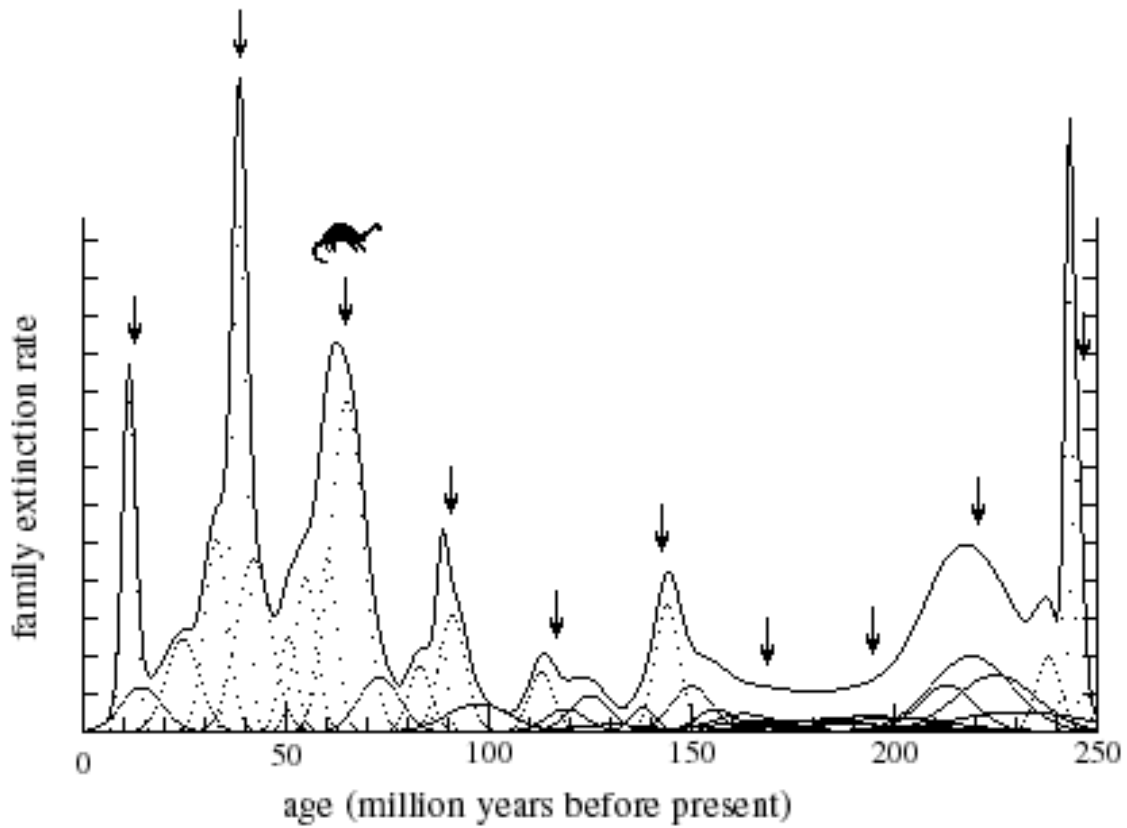 , height=12.0cm}}
\end{center}
\caption {Extinction rates versus geological time. Each data point is plotted
as a Gaussian, with width equal to the uncertainty of the geological age,
and area equal to the extinction magnitude.}
\label{Fig5}
\end{figure}

``Most discoveries in physics are made because the time is ripe'' \cite{54}.
And not only in physics. Although Fischer and Arthur had already suggested 
a 32-Myr periodicity in marine mass extinctions \cite{55}, it took about 
seven years for the subject to become popular. And this happened when 
Raup and Sepkoski's seminal paper \cite{56} appeared. They used extensive
extinction data about 3500 families of marine animals Sepkoski had collected
for years. After scrutinizing the data, only 567 families were selected for
which the data were considered as the most reliable. The extinction rates
of these families plotted versus the geological time exhibited a puzzling
periodicity. Fig.\ref{Fig5} shows Raup and Sepkoski's original data as 
presented by Muller \cite{57} 

The geological time scale accuracy is a rather subtle point \cite{58} and
not everybody agrees that the periodicity is statistically significant.
But we think that Raup and Sepkoski's analysis should be considered as at
least a strong indication of 26-30~Myr periodicity in the extinction data.
Especially if you take into account that the same periodicity was 
confirmed in Sepkoski's later studies of fossil genera \cite{59}. A similar
periodicity has been observed in the cratering rate on the Earth 
\cite{60,61},  
in magnetic reversals \cite{62} and in orogenic tectonism \cite{63}.

But if this mysterious periodicity is indeed real, you need some 
extraordinary explanation for it. Some such explanations were 
suggested shortly after Raup and Sepkoski's findings. All of them use 
extraterrestrial causes to explain terrestrial mass extinctions. This is not
surprising because only in astronomy one can find clocks with such a large
period.

Rampino and Stothers suggested \cite{61} that the Sun's motion perpendicular
to the galactic plane can modulate comet fluxes streaming towards the inner
solar system, because when the Sun crosses the galactic plane twice in his
$\sim$60~Myr period oscillations the probability to meet molecular clouds 
increases. Of course, it is an interesting fact that the half-period of 
solar oscillations perpendicular to the galactic plane practically coincides 
to the mass extinctions period. But at least two obvious drawbacks of this
hypothesis can be indicated. First of all, the present amplitude of the solar
oscillations perpendicular to the galactic plane is comparable with the scale
of molecular clouds height. So it is unlikely these Sun's oscillations to be
able to produce any detectable periodicity in encounters with molecular 
clouds \cite{64}. Besides, the Sun's oscillations in and out of the galactic
plane are out of phase with mass extinctions: the Sun is presently just near
the galactic plane, whilst we are about half-way between extinctions 
\cite{65}.

Another mechanism, which can lead to periodic comet showers, postulates 
the existence of yet undiscovered tenth planet (planet X) in the
solar system \cite{66}. It is assumed that this planet had swept out a gap
in the comet disk beyond the orbit of Neptune during its lifetime. If the
orbit of planet X has modest eccentricity and inclination to the ecliptic,
it will pass close to the inner and outer edges of the gap twice in its 
perihelion precession period. And this precession period is expected to be
about 56~Myr -- nearly twice the extinction period, if the semi-major axis
of the orbit is $\sim$ 100~AU -- big enough to ensure that it is not 
a simple matter to discover such planet. This is an interesting hypothesis 
but the question with it is whether the needed gap in the comet distribution
around the tenth planet could be maintained \cite{67}.

\begin{figure}[htb]
  \begin{center} \mbox{\epsfig{figure=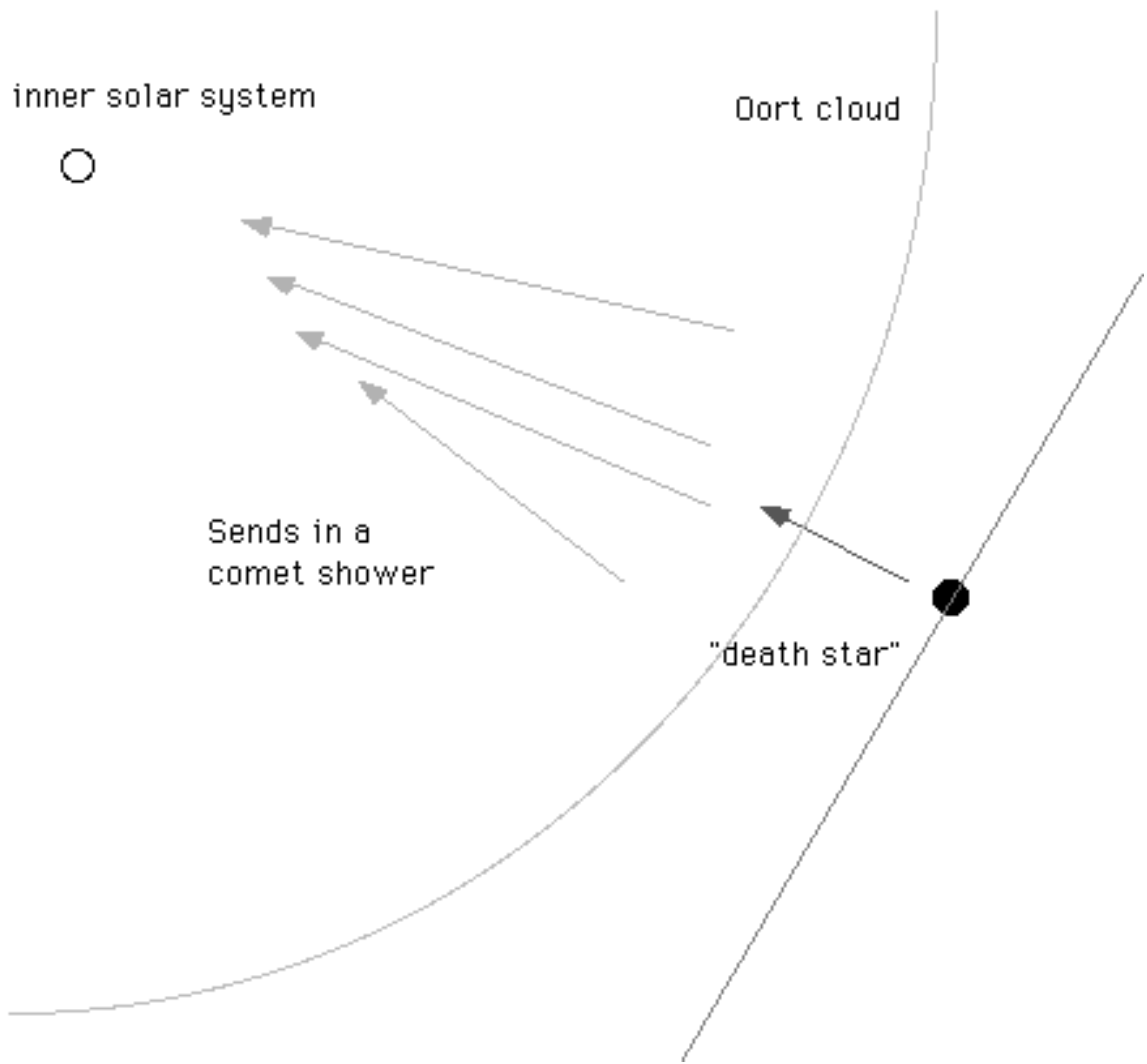 , height=14.0cm}}
\end{center}
\caption {The companion star induces comet shower while passing near the
inner Oort cloud.}
\label{Fig6}
\end{figure}

Most solar-type stars have companion(s). Partially based on this observation,
Davis et al. \cite{65} and independently Whitmire and Jackson \cite{68}
suggested that the Sun maybe is no exception and also has a distant companion
star. How can this putative solar companion cause periodic comet showers?
If its orbital period is $\sim$26~Myr it will have a large semi-major axis
$a\approx 8.8\cdot 10^4~AU\approx 1.4$ light years according to the Kepler's 
third law. But even in this case its perihelion $r_{min}=a(1-e)$, where $e$
stands for the orbital eccentricity, can be of the order of $3\cdot 10^4$~AU 
if $e\approx 0.7$, sufficiently low to disturb the inner Oort cloud -- a
comet reservoir containing about $10^{13}$ comets.  Then every perihelion 
passage of the companion star will induce a cometary shower which after
some tens of thousand years will enter the inner solar system and some of
them will hit the Earth with high probability. Schematically this is shown in
Fig.\ref{Fig6} \cite{69}.  

The hypothetical solar companion star was named Nemesis, ``after the Greek
Goddess who relentlessly persecutes the excessively rich, proud and 
powerful'' \cite{65}. This name became the most popular, although the Hindu
God of destruction Shiva and his Mother Goddess Kali were argued to be
alternatives more suitable to convey dual aspects of mass extinctions
\cite{67,70}.

Let us take a bit closer look at the Nemesis theory and estimate how many
comets are expected to hit the Earth because of the Oort cloud perturbation
caused by Nemesis. To do this, we need some model for the distribution of
comet orbits in the inner Oort cloud and we take the simplest model 
\cite{67}: all comets have the same semi-major axis $a=10^4~AU$ and their
positions and velocities are uniformly distributed in the phase space. Only
comets with the perihelion distance $a(1-e)<1~AU$ cross the Earth's orbit
and for each crossing have some chance to hit the Earth. These comets should
have orbital eccentricities $e>1-1~AU/a=1-10^{-4}$. So the fraction $\nu$
of the inner Oort cloud comets which will cross the Earth's orbit twice within 
1~Myr, the cometary orbital period for our choice of their semi-major axis,
is given by
\begin{eqnarray}
\nu=\int\limits_{0.9999}^1 f(e)de.
\label{eq11} \end{eqnarray}
\noindent Here $f(e)$ is a distribution function for the eccentricity $e$.
Because, for fixed semi-major axis, $1-e^2\sim L^2$, $L$ being the orbital 
angular momentum, the distribution function for $e^2$ is the same as the 
distribution function for $L^2$. The latter can be derived from our 
supposition about the uniform distribution of the comets in the phase space.
But it is possible to guess this distribution function more easily by using
the analogy with a highly excited quantum-mechanical hydrogen atom \cite{67}.
For highly excited states $L^2\sim l^2$, where $l\gg1$ is the total angular
momentum quantum number. Let us ask: if one excites a hydrogen atom what
is the probability that the quantum number $l$ will lay within the range
from $l$ to $l+\Delta l$? Each hydrogen atom level is $(2l+1)$-fold 
degenerate. So the desired probability will be proportional to
$$\sum\limits_l^{l+\Delta l} (2l+1)\approx \int\limits_l^{l+\Delta l} 2l
\approx 2l\Delta l,$$
\noindent where we have assumed $l\gg 1$. Therefore, the distribution function
for $l$ is $g(l)=2l$ in the classical limit $l\gg 1$. This means that
$l^2$ is distributed uniformly, and so does $L^2$ and hence $e^2$. But if
$e^2$ is distributed uniformly, the distribution function for the eccentricity
will be $f(e)=2e$ and (\ref{eq11}) gives
$$\nu=\int\limits_{1-10^{-4}}^1 2e de\approx 2\cdot 10^{-4}.$$
\noindent The total number of comets in the inner Oort cloud is estimated to
be $N=10^{13}$. Therefore $\nu N\approx 2\cdot 10^9$ comets will rush towards 
the Earth in every 1~Myr. The geometrical cross section of the Earth 
constitutes $1.8\cdot 10^{-9}$ part of its orbital area. And this number
should be even slightly enhanced because of the gravitational focusing
(about $1.1$-times \cite{67}). Therefore the expected number of comet hits on 
the Earth's surface is about $2\cdot 10^{9}\cdot 1.8\cdot 10^{-9}\cdot 1.1
\cdot 2\approx 8$. Here the last factor $2$ accounts for the fact that a comet
will cross the Earth's orbit twice during its perihelion passage and, 
therefore, will have two chances to hit the Earth.

This estimate indicates that the Earth would be a very hazardous place, hardly
capable to develop any complex forms of life, unless it has some protection
against these comet storms. And it is really protected by its faithful 
safeguards Jupiter and Saturn. Most of the comets crossing Saturn's orbit
will be ejected from the solar system after a few orbital period due to
gravitational perturbations by Jupiter and Saturn. Because of this effect,
the distribution of the Oort cloud comets in the phase space is in fact not
uniform: the region corresponding to orbits that enter the inner solar system,
the so called ``loss cone'', is normally empty. Therefore the Earth usually 
sits secure in the quiet ``eye'' of the comet storm \cite{65}.

Do you realize that we owe our opportunity to attend this conference to
Jupiter? I was quite amazed when this thought crossed my mind while preparing
these notes. Complex life might be quite rare in the universe \cite{71}. 
It is not sufficient to find a star like the sun which has a planet like 
the Earth. You need also to supply respective safeguards. 

When Nemesis comes close, it disturbs Oort cloud comets and, as a result,
fills the loss cone. In other words, this means that about two billion 
comets are sent towards the Earth each time Nemeses passes its perihelion.
The total number of impacts expected on Earth will be higher than eight --
our above estimate. Paradoxically, this is due to effects of Jupiter and 
Saturn. A small number of comets from the Nemesis induced shower will not
be immediately expelled from the solar system by these safeguards but
instead perturbed into smaller, frequently returning orbits. This comets
will visit the planetary system several times until their final ejection
on hyperbolic orbits or disintegration due to a close approach to the Sun.
Hence the probability to hit the Earth increases several times, up to order
of magnitude \cite{67}.

As we see, if the Nemesis is heavy enough to fill the loss cone, its close
approaches to the Sun will be catastrophic for creatures like dinosaurs.
Smaller creatures, like cockroaches, can possibly survive and enjoy the night
sky filled with comets, with several new comets appearing every day. It was 
shown \cite{67} that if the mass of the Nemesis is not much smaller than
$0.1~M_\odot$, the loss cone will be indeed filled by a single perihelion
passage of the perilous solar companion for assumed eccentricity $e=0.7$.

The next obvious question to be answered before acceptance of the Nemesis 
theory is the stability of such a wide binary system. While orbiting the
Sun, Nemesis experiences both slowly changing and rapidly fluctuating
perturbations. The former is due to galactic tides and the Coriolis forces
(remember that the solar rest frame rotates around the galactic center).
The latter is caused by passing field stars and interstellar clouds.

For assumed semi-major axis, the Nemesis is in the region where the Sun's
gravity still dominates over the Galaxy field. But due to galactic tides,
the orbit oriented parallel to the galactic plane is more stable than orbits
at higher galactic latitudes \cite{67,72,73}. Moreover, retrograde orbits
are more stable because for such orbits Coriolis forces increase stability
\cite{73}. Therefore it may be more probable for the Nemesis to be located
at low inclinations with respect to the galactic plane. But it is not 
excluded that present day Nemesis has high inclination, because its orbit
is not rigid but subject to various perturbations. So one can imagine that
Nemesis started with low inclination and much less wide orbit and random
perturbations had lead to its present wide and high galactic latitude orbit,
where it can still have several hundred Myr lifetime \cite{67} (according
to \cite{73}, the lifetime for an orbit perpendicular to the galactic plane 
is $\sim$500~Myr).

The perturbing effects of passing field stars were studied by extensive
numerical calculations \cite{72,74}. It was shown that the period of 
``double star clock'' fluctuates randomly due to this effect. But the
expected drift in orbital period over last 250~Myr (the geological period
of interest in a light of Raup and Sepkoski's data) is within a 10 to 20 \%
-- low enough not to spoil periodicities in observable mass extinction data.

The lifetime of $10^3$~Myr for the Sun-Nemesis system found in this 
calculations suggests that it is not possible for the Nemesis to be on such
wide and eccentric orbit all the time during solar system existence. So 
either Nemesis was captured by the Sun relatively recently -- the event
considered as extremely unlikely \cite{75} because it requires three-body
encounters or very close encounters to allow a tidal dissipation of the
excessive energy, or its orbit was much more tight at early years of solar
system and random-walked to its present position. In the latter case one
can expect higher bombardment rate in the past. And it is known that at least
in the period between 4.5 and 3~Gyr the bombardment rate was indeed very
high. It is believed that one such collision of a planetary size object with 
the Earth lead to the formation of the Moon. Intriguingly, a moon of right
size and at right position appears to be one more ingredient for complex life 
to develop on the Earth's surface \cite{71}, because it minimizes changes in
the Earth's tilt, ensuring climate stability.

One more important question was successfully settled by these numerical
calculations. In principle, some perturbation can force Nemesis to enter into
the planetary system and cause ``a catastrophe of truly cosmogonical 
proportions'' \cite{74}. Fortunately, this fatal event turned out to have 
a very low probability and hence the planetary system can survive the presence
of distant solar companion \cite{72,74}.

The effects of interstellar clouds are the most uncertain. Opinions about the
fate of Sun-Nemesis system here change from extreme pessimistic \cite{76} to
extreme optimistic \cite{77}. The truth should lay somewhere in the middle
between these two extremes. Unlike a field star, a single close encounter
with a giant molecular cloud can instantly disrupt a wide binary. But in 
contrast to the stellar neighborhood of the Sun, both the distribution and 
internal structure of the interstellar clouds are poorly known near the Sun
\cite{67}. Disruptive effects of interstellar clouds were investigated by
Hut and Tremaine \cite{78}. Their analysis indicate that the effects of 
interstellar clouds lead most probably to the lifetime of $10^3$~Myr for
distant solar companion, comparable to the lifetime caused by stellar
perturbations \cite{67}. Therefore interstellar clouds seem to be also harmless
for the Nemesis hypothesis if Nemesis begins its career at much tighter orbit
than the postulated present orbit.

To summarize, there are some indications of 26-30~Myr periodicity in mass
extinction data and in some other geological phenomena. This periodicity can
naturally explained if we assume existence of a distant solar companion star
-- Nemesis. Its present orbit is not stable enough to ensure such a wide and
eccentric orbit all the time since the solar system formation. But if the 
Nemesis was on a much more tighter orbit in the past and random-walked to its
present position due to various perturbations, nothing seems to invalidate the
hypothesis. The only drawback of this theory is that Nemesis was never found.
And this is the point where mirror world enters the game: you can't expect to
discover Nemesis through conventional observations if it is made from some 
mirror stuff, can you?

But why mirror Nemesis? Is any more serious reason for the God, except to hide
Nemesis from us, to choose the mirror option? Maybe there is. While looking at
the solar system, an obsessive impression appears that every detail of it was
designed to make an emergence of complex life possible \cite{71}. And it took
billions of years of evolution for creatures to appear as intelligent as we 
are. Nemesis, it is believed, had played an important role in this process,
periodically punctuating evolution. Therefore you need the Nemesis to orbit
for ages. As we have mentioned earlier, the best way to do this is to place
the Nemesis orbit at lower galactic latitudes to minimize disruptive effects
of galactic tides and hence increase the orbit stability. But if Nemesis is
made from the ordinary matter and was formed from the same nebula as the rest
of the solar system, you expect the Nemesis orbit to be in the ecliptic plane 
-- at high galactic latitudes. On the contrary, if the solar system was formed
from a nebula of mixed mirrority (the possibility of such nebula was 
considered in \cite{79}), a priori there is no reason the mirror part of the
nebula to have the same angular momentum direction as the ordinary part. So 
for mirror Nemesis it is natural just to be formed in a plane different from
the ecliptic.

Of course, the above given arguments are not completely rigorous. But who 
knows, maybe the answer on the question ``what killed the dinosaurs?'' really
sounds like this: Nemesis, the mirror matter sun.

\section{Conclusions}
The mirror Nemesis hypothesis emerged almost as a joke during our e-mail
discussions with Robert Foot. After some thought we found no reason why this
hypothesis, although somewhat extravagant, might not be true \cite{80}.
As a result, the dinosaur theme, which I originally intended to introduce
for just to make presentation more vivid, quickly became one of the central
motives of this talk, and you have the story presented above. I hope you
enjoyed it regardless whether dinosaurs were really eyewitnesses of the 
mirror world or not.

I consider the possibility to restore the equivalence between left and right
through the mirror world as very attractive. Theories with extra spatial
dimensions, and $M$-theory in particular, can easily produce various ``shadow
worlds'' which are however not necessarily parity invariant (this refers to
the $E_8\times E_8$ model also, mentioned earlier), but some of them might 
be, so realizing the mirror world scenario. Maybe, it is even possible to 
have a mirror world without mirror particles. $M$-theory nonorientable
compactifications, suggested so far \cite{81}, do not lead to the realistic 
model, as I can judge. But it will be very interesting to find a realistic
example and show that the parity noninvariance of our world indicates to its
nonorientable topology and is only local phenomenon.

``In the Soviet scientific society the scientists had one freedom that 
scientists in the West lacked and still lack (perhaps the only real freedom 
that Eastern scientists had), and that was to spend time also on esoteric 
questions. They did not have to be scrutinized by funding agencies every now 
and then'' \cite{82}. The Soviet Union disappeared and so did this freedom. 
You can consider this paper, if you like, as a nostalgia for this
kind of freedom, enabling to escape bonds of the stiff pragmatic logic. 

\section*{Acknowledgements}
Although, as I became aware while preparing these notes, historically I owe
my chance to attend this beautiful place and conference to Jupiter, the Moon
and Nemesis, all their efforts would be in vain without professor Zurab
Berezhiani. I thank him very much for giving me a possibility to attend the
Gran Sasso Summer Institute, and for his kind hospitality during the
conference. I also thank to Denis Comelli, Francesco Villante and Anna Rossi
for their help at Ferrara and Gran Sasso.

I'm indebted to Sergei Blinnikov for encouragement and for indicating the
Hubble-Hipparchos controversy.

The content of this talk would be very different unless fruitful discussions
with Robert Foot, which I acknowledge with gratitude.

I thank Piet Hut for sending me reprints of his articles, which were 
heavily used in these notes.

\end{document}